\documentclass[12pt]{article}%
\usepackage[T2A]{fontenc}
\usepackage[cp1251]{inputenc}
\usepackage{amsmath}
\usepackage{amsfonts}
\usepackage{amssymb}
\usepackage{graphicx}
\usepackage{cite}%
\usepackage{authblk}

\begin{document}

\title{Phase space representation of sound field in the Lake Kinneret}
\date{\ }

\author[1]{A.L. Virovlyansky}
\author[1]{A.Yu. Kazarova}
\author[2]{B.G. Katsnelson}
\affil[1]{\footnotesize Institute of Applied Physics, Russian Academy of Science, 46 Ul'yanov Street, Nizhny Novgorod, 603950 Russia, Virovlyansky@mail.ru}
\affil[2]{\footnotesize L.Charney School of Marine Sciences, University of Haifa, 199 Aba Khoushy Ave, Haifa, 3498838 Israel, bkatsnels@univ.haifa.ac.il}

\maketitle

\begin{abstract}
The paper presents the analysis of pulsed sound fields recorded by a vertical array in the Lake  Kinneret (Israel). The transition from the traditional representation of the complex amplitude of the received field as a function of depth and time to a function representing the field distribution in the phase space `depth - angle - time' is considered. Due to the absence of multipath and problems with caustics, the sound field distribution in phase space is less sensitive to environmental disturbances and therefore more predictable than in configuration space. The transition is carried out using the coherent state expansion developed in the quantum theory. The found distribution of the field intensity in the phase space agrees with the calculation performed with an idealized environmental model. It is shown that this distribution can be taken as the input for solving the problem of source localization. The results of data processing demonstrate the possibility of using the coherent state expansion for isolating the field component formed by a given beam of rays.
\end{abstract}

\section{Introduction}
A characteristic feature of sound propagation in
underwater waveguides is multipath \cite{JKPS2011,Etter2018}. The
field at the observation point usually represents a superposition
of signals coming along different paths and therefore crossing
different random inhomogeneities that are not taken into account
in the available environmental model. As a result, the calculation
of the complex amplitudes of these signals often turns out to be
insufficiently accurate for theoretical prediction of the total
field. This problem, which significantly complicates the solution
of almost all problems of underwater acoustics, (source
localization, underwater communication, remote sensing, etc.) is
referred to as the uncertain environment
\cite{Livingston2006,Dosso2016}.

The present paper considers the approach introduced in Refs.
\cite{V2017,V2020a} for analyzing the sound field under condition
of uncertain environment. This approach is based on the transition
from the traditional description of the complex amplitude of the
tonal field in the vertical section of the waveguide as a function
of the depth $z$ to the amplitude distribution in the phase plane
`depth - grazing angle'. This transition is carried out using the
coherent state expansion developed in quantum mechanics
\cite{Glauber2007,Klauder,Shl2001}. The distributions of the
amplitude and, especially, the intensity (squared amplitude) of
the sound field in the phase space, represented in this example by
the phase plane, turn out to be less sensitive to the medium
perturbation than their distributions in the configuration space,
represented by the $z$ axis. The reason is that there is no
multipath in phase space: no more than one ray trajectory comes to
any point in this space \cite{Gold2000}. Similarly, when analyzing
a field excited by a pulsed source, we can pass from the
distribution of the field amplitude in the 2D phase plane `depth -
time' to its distribution in the 3D phase space `depth -- angle --
time'.

The function expressing the complex field amplitude dependence on
the phase space coordinates is called the phase space
representation of the sound field. Similar representations of wave
fields are used in optics \cite{Shl2001,Alonso2010}.

Since the field amplitude distribution in the phase space, as
noted above, is more stable with respect to the medium
perturbation than the amplitude distribution in the configuration
space, the transition to the phase space representation can reduce
the requirements for the accuracy of environmental model when
solving inverse problems. In Refs. \cite{V2020a,V2020} this is
demonstrated by the example of solving the problem of source
localization in a waveguide.

In the present paper we examine the phase space representation  of
a pulsed sound field from a point source recorded in the Lake
Kinneret (Israel) by a vertical receiving array. The measurements
were taken in 2019 and 2021. The main attention is paid to the
comparison of the received field intensity distribution in the
phase space `depth - angle - time' with the results of numerical
simulation.

The paper is organized as follows. Section \ref{sec:experiment}
describes the experiments and the data obtained. Two idealized
models of range-independent waveguide used for numerical
simulation are also presented here. The main provisions of the
theory used in the construction of the phase space representation
are outlined in Sec. \ref{sec:theory}. The results of data
processing are presented in Sec. \ref{sec:data}. In this section,
the calculated and measured distributions of the field intensity
in the phase space 'depth - angle - time' are compared. This
section also demonstrates the possibility of using these
distributions as input to solve the source localization problem.
The results of the work are summarized in Sec.
\ref{sec:conclusion}.

\section{Experimental data and environmental model \label{sec:experiment}}

Subtropical Lake Kinneret (the Sea of Galilee) is an example of a
fresh water reservoir, with changing stratification: approximately
constant temperature 15-16$^\circ$ C  over all the depth in the
Winter and reaching 30$^\circ$ C  in upper layer in the Summer.
Approximate size is 12 x 22 km, maximal depth of the lake is 40 m,
depth of thermocline 10-20 m.

Acoustic measurements were carried out in 2019 and 2021 in the
central part of the lake where the bottom is almost flat. The
signals were recorded using a receiving vertical array of 10
hydrophones, and the radiation was performed by a source lowered
from a drifting vessel. The table shows the distances from which
the signals were recorded.

\begin{table}[!t]
\caption{Observation distances} \centering
\begin{tabular}{|c||c|}
\hline
\text{Year} & \text{Distances}\\
\hline
\text{2019} & \text{340 m, 380 m}\\
\hline
\text{2021} & \text{415 m, 905 m, 1445 m}\\
\hline
\end{tabular}
\end{table}

In 2019, at each of the indicated distances, a source located at a
depth of 10 m emitted chirp pulses with a duration of 1 s in the
frequency band from 300 to 3500 Hz. The receiving hydrophones
covered the depth interval from 10 to 37 m with a step of 3 m. The
sound speed profile measured near the receiving array is shown on
the left side of Fig. 1. The dots show the depths of the receiving
hydrophones.

In 2021, at the distances indicated in the table, the source was
located at a depth of 7 m and emitted chirp pulses with a duration
of 5 s in the frequency band from 200 Hz to 10 kHz. The sound
speed profile and hydrophone depths are shown on the right side of
Fig. 1. The hydrophones covered the depth interval from 7 to 34 m
with a step of 3 m.

\begin{figure}[!t]
\centering
\includegraphics[width=4.5in]{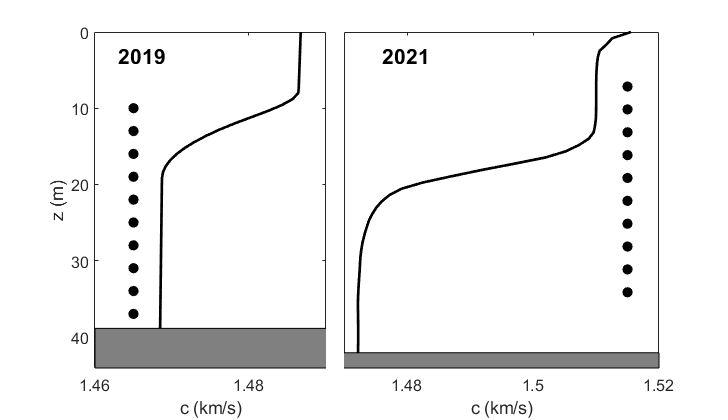}
\caption{Sound speed profiles and hydrophone positions in 2019
(left panel) and 2021 (right panel).} \label{fig_1}
\end{figure}

When simulating the measured acoustic fields, we used
range-independent waveguide models with sound speed profiles from
the left and right sides of Fig. 1. Due to the difference in water
surface levels, the waveguide depths in 2019 and 2021 were
slightly different (this is shown in Fig. 1). In simulation, they
were taken equal to 38.9 m and 42.1 m, respectively.

The bottom structure of Lake Kinneret is complex.  In central
area, the sediment is composed mainly of clays and carbonate. The
presence of gas (methane) bubbles in the upper sediment was found
earlier and confirmed by many direct measurements. Both
concentration of bubbles (less than 0.5 - 1\%) and thickness of
gassy layer (less than 50 - 70 cm) change in dependence on place
and season \cite{Katsnelson}. For our purpose in this paper, we
can use a simplified geoacoustic model representing a liquid
homogeneous half-space.

In the present paper, our objective is to analyze the field
components formed by narrow beams of rays whose grazing angles
near the bottom do not exceed $42^{\circ}$. In Lake Kinneret, due
to the high concentration of gas bubbles in sediments, waves with
such grazing angles are almost completely reflected from the
bottom. In both our waveguide models, the bottom is represented by
a liquid half-space with a sound speed of 2 km/s and a density of
1400 kg/m$^{3}$. Although these environmental models are obviously
inexact, they correctly reflect the fact that waves with the
indicated grazing angles are completely reflected from the lower
boundary of the waveguide. In this case the reflection
coefficients for all rays of a narrow beam are approximately equal
to $V = \exp(i\phi)$, where $\phi $ is the angle whose value
cannot be predicted within the framework of our model. We assume
that, despite the inaccuracy of the bottom model, the desired
field components can be approximately calculated up to unknown
phase factors $V^N$, where $N$ is the number of beam reflections
from the bottom. Below we will see that this is sufficient for the
evaluation of the sound field intensity in the phase space.

The sound field of a point source in a range-independent waveguide
is a function of distance $r$, depth $z$, and time $t$. The $z$
axis is directed vertically downwards and the water surface is at
the horizon $z=0$. At the observation distance, that is, for a
fixed $r$, we represent the complex field amplitude as the Fourier
integral
\begin{equation}
v\left(  z,t\right)  =\int df~u\left(z,f\right)  e^{-2\pi ift}%
.\label{v-u-0}%
\end{equation}
The calculation of the Fourier components $u\left( z,f\right) $ at
frequencies $f$ in the band of the emitted signal was performed by
the normal mode method \cite{JKPS2011,BL2003}.

As mentioned above, the sound field $v\left( z,t\right) $ was
measured only at 10 horizons $z_{n}$, $n=1,\ldots,10$.
Accordingly, the values of functions $u\left( z,f\right) $ are
known only at these depths.Using these data, one can approximately
estimate the amplitudes of the propagating modes
\cite{Morozov2010} and then reconstruct functions $u\left(
z,f\right) $ over the entire vertical section of the waveguide.
Let us use the representation of the function $u\left( z,f\right)
$ as a superposition of eigenfunctions $\varphi_{m}\left(
z,f\right) $ of the Sturm-Liouville problem for a model waveguide

\begin{equation}
u\left(  z,f\right)  =\sum_{m=1}^{M\left(  f\right)  }b_{m}\left(
f\right)
\varphi_{m}\left(  z,f\right)  , \label{u-phi}%
\end{equation}
where $M\left( f\right) $ is the number of propagating modes.
Substituting the values of functions $u\left( z,f\right) $ and
$\varphi_{m}\left(z,f\right) $ at 10 points $z_{n}$ into this
equality, we get (at each frequency $f $)  a system of 10
equations for the $M\left( f\right) $ unknowns $b_{m}\left(
f\right) $. At all frequencies in the band of the emitted signal,
this system is underdetermined. Its approximate solution can be
found using the pseudoinverse matrix. Substituting the found
values $b_{m}\left( f\right) $ into (\ref{u-phi}) gives an
expression for calculating the field at an arbitrary depth. Due to
the fact that there are only ten receiving hydrophones and the
distance between neighboring hydrophones (3 m) is relatively
large, the obtained solution of the inverse problem has an
acceptable accuracy only at low frequencies in the $300$ Hz
$<f<900$ Hz band.

To eliminate the contributions of higher frequencies, the recorded
signals were bandpass filtered with the spectral weight
$\exp\left( -\pi\left(f-f_{c}\right) ^{2}/\Delta_{f}^{2} \right)
$, where $f_{c}$ = 600 Hz and $\Delta_{f}$ = 300 Hz. The
amplitudes of bandpassed signals at the horizons $z_{n}$, received
from  distances of 380 m and 905 m, are shown in Figs. 2 and 3,
respectively. Here and in what follows, for brevity, we indicate
only the distance to the source, omitting the year of measurements
(see Table I). The upper and lower panels of the figures show the
results of theoretical calculations and the measurement data,
respectively. In each plot, the magenta broken lines represent the
timefront depicting ray arrivals in the $\left( t,z\right) $
plane. The timefronts were calculated using the corresponding
waveguide models. The figures show the initial sections of the
recorded signals with a duration of 0.08 s. Further, when
comparing theory and experiment, we analyze fragments of the
received fields in this time interval.

\begin{figure}[!t]
\centering
\includegraphics[width=4.5in]{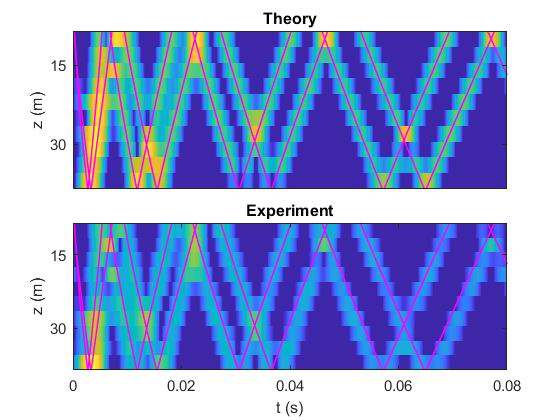}
\caption{Signal amplitudes $\left\vert v\left( z,t\right)
\right\vert $ on receiving array elements recorded from a distance
of 380 m. Top: numerical simulation. Bottom: measurement data. The
magenta broken lines represent the time front.} \label{fig_2}
\end{figure}

\begin{figure}[!t]
\centering
\includegraphics[width=4.5in]{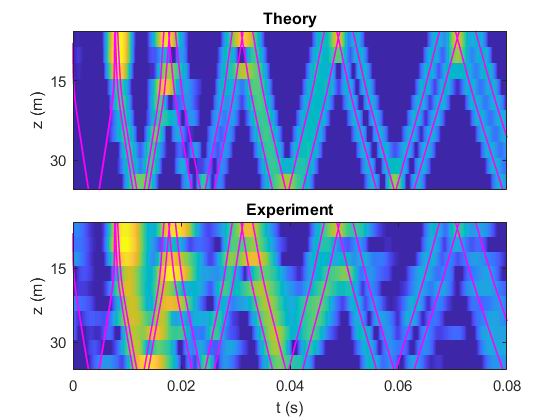}
\caption{The same as in Fig. 2, but for signals received from 905
m.} \label{fig_3}
\end{figure}

\section{Theory \label{sec:theory}}

In this section, we consider a theoretical description of the
field excited by a point source in a range-independent waveguide
with a sound speed profile $c \left( z\right) $ and a refractive
index $n\left( z\right) =c_{0}/c\left( z \right) $, where $c_{0}$
is the reference sound speed.

\subsection{Ray paths in the phase space \label{sub:paths}}

The phase space appears in the Hamiltonian formulation of
classical mechanics and geometrical optics
\cite{Gold2000,Alonso2010,Vbook2010}. Within the framework of this
formalism, the ray trajectory at distance $r$ is determined by its
depth $z$ and momentum $p=n\left( z\right) \sin\chi$, where $\chi$
is the grazing angle at the point $\left (r,z\right) $. The ray
equations take the form of the Hamilton equations $dz/dr=-\partial
H/\partial p$, $dp/dr=-\partial H/\partial z$, where
$H=-\sqrt{n^{2}\left( z\right) -p^{2}}$ is the Hamiltonian. The
functions $p\left( r,p_{0},z_{0}\right) $ and
$z\left(r,p_{0},z_{0}\right) $ representing solutions of these
equations with initial conditions $p\left( 0\right) =p_{0}$ and
$z\left( 0\right) =z_ {0}$, describe the ray trajectory.

The ray travel time $t\left( r,p_{0},z_{0}\right) $, i.e. the
travel time of sound along the ray path, is analogous to the
action integral or the Hamilton's principal function in mechanics.
This function is given by the integral along the ray path
$c_{0}t=\int\left( pdz-Hdt\right)$ \cite{Alonso2010,Vbook2010}.

In the case of a point source, all rays at $\ r=0$ escape from the
same depth $z_{s}$ with different launch angles $\chi_{0}$ and,
accordingly, with different starting momenta $p_ {0}=n\left(
z_{s}\right) \sin\chi_{0}$. The arrival of a ray at a given
observation distance $r > 0$ is represented by a point in the
phase space $\left( z,p,t\right) $. The set of such points forms a
curve, which we call the \textbf{ray line } in the 3D phase space
$\left(z,p,t\right) $. It is parametrically determined by the
equations: $p= p\left(r,p_{0},z_{s}\right) $, $z=z\left(
r,p_{0},z_{s}\right) $ and $t=t\left(r,p_{0},z_{s}\right) $.

Choosing the reference sound speed $c_{0}$ = 1.5 km/s in
waveguides with $c\left( z\right) $ profiles shown in Fig. 1, we
get $n$ values close to one. For flat rays, the momentum $p$ is
approximately equal to the grazing angle $\chi$. Therefore, the
phase space $\left( z,p,t\right) $ can be called the 'depth --
angle -- time' space.

\begin{figure}[!t]
\centering
\includegraphics[width=4.5in]{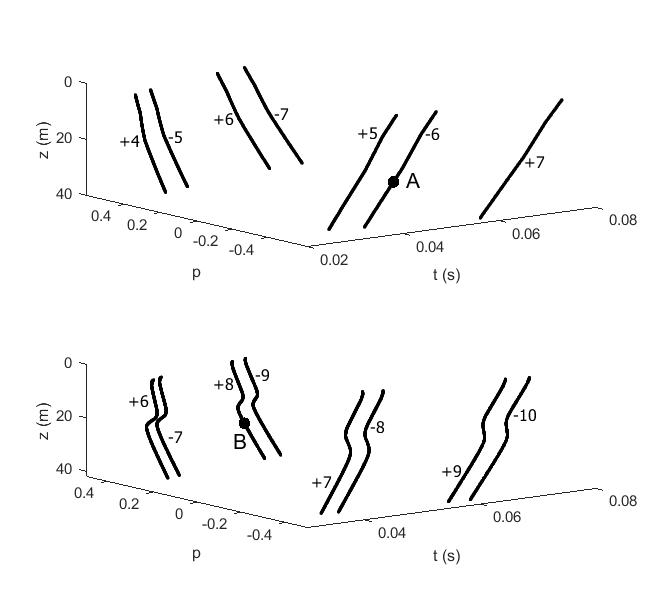}
\caption{Fragments of ray lines at a distance of 380 m in the time
interval of 0.02$\div$0.07 s (upper panel) and at a distance of
905 m in the time interval of 0.03$\div$0.07 s (lower panel). The
black circles $A$ (top panel) and $B$ (bottom panel) show the
arrivals of rays that escape the source with launch angles
$-30^{\circ}$ and $19.5^{\circ}$, respectively.} \label{fig_4}
\end{figure}

Figure 4 shows fragments of ray lines at distances of 380 m (upper
panel) and 905 m (lower panel). The ray line consists of segments
formed by trajectories with the same identifiers $\pm M$, where
$\pm$ is the sign of launch angle (+ corresponds to the rays
starting towards the bottom), and $M$ is the number of turning
points. In our case, for almost all rays, the turning points are
the points of reflection from the boundaries. In what follows, the
segment depicting the arrivals of rays with the identifier $\pm M$
will be called, for brevity, the segment $\pm M$.

Note that for rays propagating without reflections from boundaries
in a refracting waveguide, the ray line is continuous
\cite{V2017,V2020a,V2020}. In our example, discontinuities arise
due to the fact that when reflected from the boundary, the ray
grazing angle, and hence its momentum, changes sign.

On each of the ray lines shown in Fig. 4, the arrival of one of
the rays is marked with a black circle. On the top panel, this is
ray $A$ which escaped the source at a launch angle of
$-30^{\circ}$. Its identifier at the observation range is -6. The
trajectory of this ray (bold line) and the trajectories of other
rays forming segment -6 are shown at the top of Fig. 5.

On the bottom panel of Fig. 4, the black circle marks the arrival
of ray $B$ whose launch angle is $19.5^{\circ}$. At a distance of
905 m, its identifier is +8. The beam of rays forming segment  +8
is shown at the bottom of Fig. 5. The trajectory of ray $B$ is
marked with a bold line.

\begin{figure}[!t]
\centering
\includegraphics[width=4.5in]{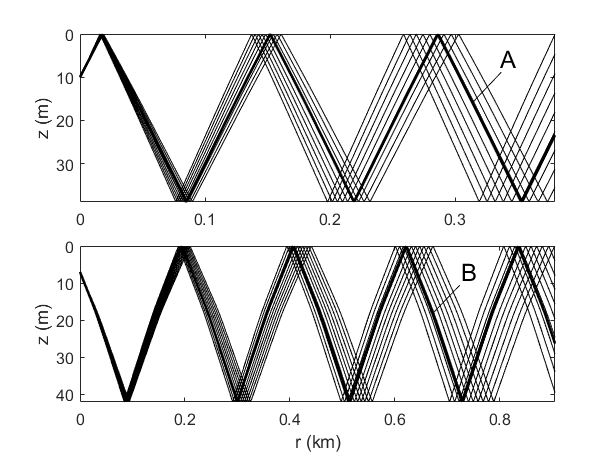}
\caption{Top panel. Ray trajectories forming segment -6 (thin
lines) at 380 m. The bold line shows the trajectory of ray $A$.
Bottom panel. Ray trajectories forming segment +8 (thin lines) at
905 m. The bold line shows the trajectory of ray $B$.}
\label{fig_5}
\end{figure}

Figures 6 and 7 show the projections of the ray lines from Figs. 4
on the planes $\left(p,z\right) $ (top), $\left( t,z\right) $
(middle), and $\left( t,p\right) $ (bottom). Each segment of the
ray line is represented by its projections onto the indicated
planes. A corresponding identifier is indicated next to each
projection. Projections onto the $\left( t,z\right) $ plane shown
in the upper panels of Fig. 6 and 7 represent the timefronts shown
by the magenta lines in Fig. 2 and 3, respectively. The arrival of
ray $A$ in fig. 6 is shown with a black circle. In addition, in
the middle panel of Fig. 6, the arrival of ray $A'$ with the
travel time equal to the travel time of ray $A$ is highlighted.
Similarly, in Fig. 7, the arrivals of ray $B$ are highlighted. The
middle panel also highlights the arrival of ray $B'$ with the same
travel time as that of ray $B$.

In the upper panels of Fig. 6 and 7, the areas representing the
so-called fuzzy segments are shown in grey. They will be
considered further in Sec. \ref{sub:isolation}.

\begin{figure}[!t]
\centering
\includegraphics[width=4.5in]{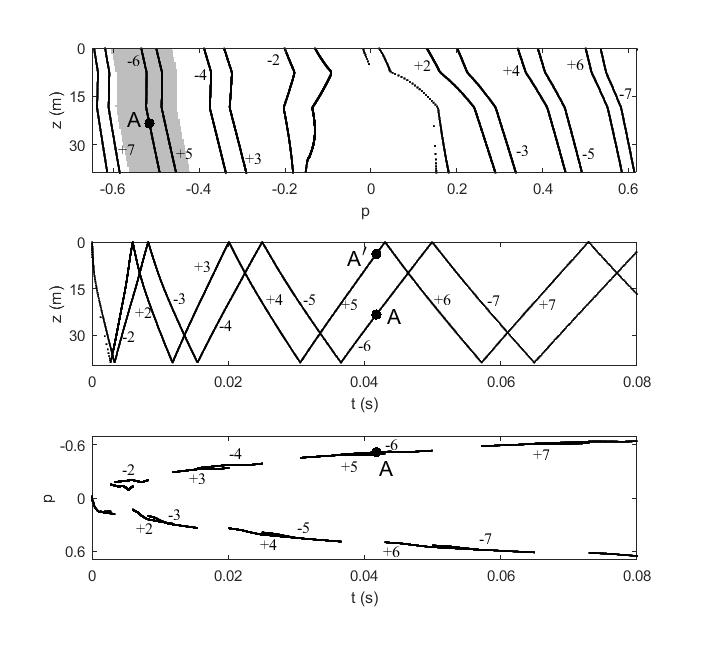}
\caption{Projections of the ray line at a distance of 380 m (Fig.
4, top panel) on the planes $\left( p,z\right) $ (top), $\left(
t,z\right) $ (middle), and $ (t,p)$ (bottom). Near each segment,
the corresponding identifier is indicated. The black circles on
all panels highlight the arrival of ray $A$. The middle panel also
highlights the arrival of ray $A'$ with the same travel time as
that of ray $A$.} \label{fig_6}
\end{figure}

\begin{figure}[!t]
\centering
\includegraphics[width=4.5in]{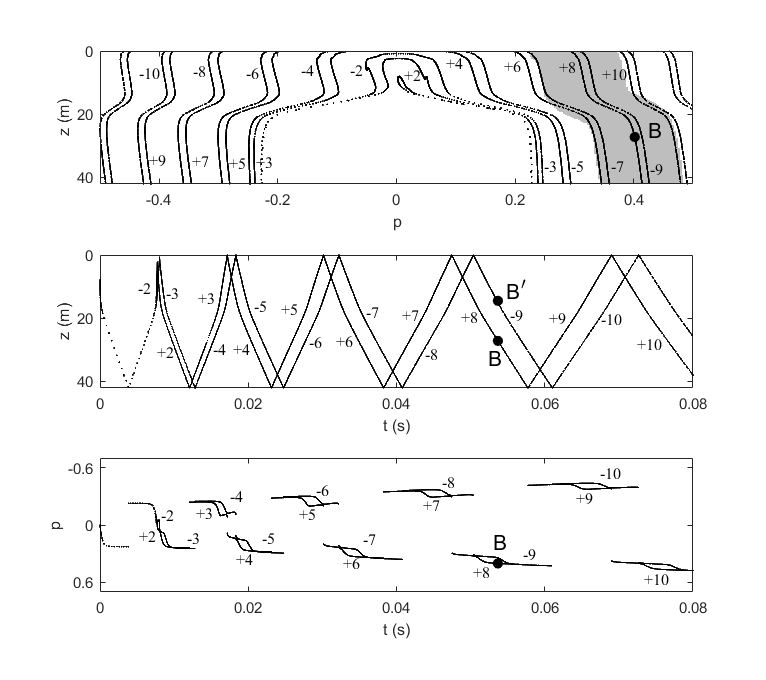}
\caption{The same as in fig. 6, but for a distance of 905 m. The
arrivals of ray $B$ and ray $B'$ whose travel time is the same as
that of ray $B$ are marked with the black circles.} \label{fig_7}
\end{figure}

\subsection{Coherent state expansion \label{sub:CS}}

The function $u\left(z,f\right)$ on the right side of
(\ref{v-u-0}) represents the depth dependence of the complex field
amplitude at some fixed frequency $f$. For brevity, in this and
the next section, the argument $f$ will be omitted. All our
subsequent analysis is based on the transition from
$u\left(z\right) $ to a function characterizing the distribution
of the field amplitude in the phase plane 'depth -- momentum
(angle)' $\left( z,p\right) $. The transition is carried out using
the coherent state expansion \cite{Shl2001}.

The coherent state associated with the point $\mu=\left(
z,p\right) $ of the phase plane is determined by the function
\begin{equation}
Y_{\mu}\left( z'\right)  =\frac{1}{\sqrt{\Delta_{z}}}\exp\left[
ikp\left( z'-z\right)  -\frac{\pi\left(  z'-z\right)
^{2}}{2\Delta_{z}^{2}}\right]  ,
\label{Y-mu}%
\end{equation}
where $\Delta_{z}$ is the vertical scale, $k=2\pi f/c_{0}$ is the
reference wavenumber. In quantum mechanics, (\ref{Y-mu}) describes
a state with the minimum uncertainty, that is, with the minimum
product of the standard deviations of the position and momentum
\cite{LLquant}. In acoustics (\ref{Y-mu}) describes the vertical
section of a wave beam with the smallest possible product of the
beam width by the spread of the grazing angles of the waves
forming it.

Although the coherent states are not orthogonal, they form a
complete system of functions and an arbitrary function $u(z')$ can
be represented as an
expansion \cite{Klauder,Shl2001}%
\begin{equation}
u\left(  z'\right)  =\lambda^{-1}\int d\mu~a_{\mu}Y_{\mu}\left(
z'\right)
,\label{u-Y}%
\end{equation}
where%
\begin{equation}
a_{\mu}=\int dz'~u\left(  z'\right)  Y_{\mu}^{\ast}\left(
z'\right)
,\label{a-Y}%
\end{equation}
superscript * means complex conjugate. In these relations,
integration over $\mu$ goes over the entire phase plane, and
integration over $z'$ goes over the entire vertical axis.

The complex amplitude $a_{\mu}$ at the point $\mu=\left(
z,p\right) $ represents the projection of the field in the
vertical section of the waveguide onto the coherent state
(\ref{Y-mu}). It quantitatively characterizes the contribution of
waves coming to depths close to $z$ at grazing angles close to
$\arcsin p$. Thus, $a_{\mu}$ characterizes the distribution of the
field amplitude in the phase plane 'depth -- angle'. The squared
amplitude $\left\vert a_{\mu}\right\vert ^{2}$ in quantum theory
is called the Husimi function \cite{Shl2001}. In what follows we
will call it the intensity of the coherent state.

The closeness of the coherent states associated with the points of
the phase plane $\mu=\left(  z,p\right)  $ and $\mu_{1}=\left(
z_{1},p_{1}\right)  $ can be quantitatively characterized by their
squared scalar product
\begin{equation}
\left\vert \int dz^{\prime}~Y_{\mu}\left(  z^{\prime}\right)  Y_{\mu_{1}%
}^{\ast}\left(  z^{\prime}\right)  \right\vert
^{2}=e^{-\frac{1}{2}d\left(
\mu,\mu_{1}\right)  },\label{Y-scalar}%
\end{equation}
where%
\begin{equation}
d\left(  \mu,\mu_{1}\right)  =\frac{\pi\left(  z-z_{1}\right)  ^{2}}%
{\Delta_{z}^{2}}+\frac{\pi(p-p_{1})^{2}}{\Delta_{p}^{2}},\label{d}%
\end{equation}
$\Delta_{p}=\lambda/(2\Delta_{z})$, $\lambda=2\pi/k$ is the
wavelength. We will interpret the quantity $d\left(
\mu,\mu_{1}\right) $ as a dimensionless distance between the
points $\mu$ and $\mu_{1}$. The coherent states associated with
these points will be considered close for $d<1$ and different for
$d>1$. The distance from the point $\mu$ to a curve in the phase
plane (for example, to the ray line or to its segment) is the
distance from $\mu$ to the nearest point of the curve.

The complex amplitude $a_{\mu}$ is defined for any point $\mu$ of
the phase plane. However, according to (\ref{a-Y}) -- (\ref{d}),
the coherent state intensities  $\left\vert a_{\mu}\right\vert
^{2}$ take on maximum values near the points corresponding to ray
arrivals. These points lie on the projection of the ray line on
the plane $\left(z,p\right) $. We will call this projection a ray
line in the phase plane. In mechanics and geometric optics, it is
called the Lagrangian manifold \cite{Alonso2010}. Examples of such
lines are presented in the upper panels of Figs. 6 and 7.

The main contribution to $u\left( z\right) $ comes from coherent
states associated with points $\mu$ located at distances $d<1$
from the ray line. We call this part of the phase plane the
\textbf{fuzzy ray line}. Its area is determined by the choice of
the coherent state scales $\Delta_{z}$ and $\Delta_{p}$. Since
they are related by the uncertainty relation
$\Delta_{z}\Delta_{p}=\lambda/2$, then in fact we are talking
about choosing one of these scales. Ref. \cite{V2020a} discusses
the choice of $\Delta_{z}$ to minimize the area. It is clear that
for $\Delta_{z}\rightarrow0$\ and $\Delta_{z}\rightarrow\infty$
the area increases indefinitely. The minimum corresponds to some
finite $\Delta_{z}$, which is determined by the shape of the ray
line. In Ref. \cite{V2020a} it is shown that this scale is
proportional to $\lambda^{1/2}$.

\subsection{Isolation of contribution from a given beam of rays to the total field 
\label{sub:isolation}}

Let us take a particular segment of the ray line in the phase
plane. In accordance with the above, the contribution to the total
field of rays represented by this segment is expressed as a
superposition of coherent states associated with the phase plane
points located at distances $d<1$ from the segment. The area
$\sigma$ formed by these points is called the \textbf{fuzzy
segment} of the phase plane. Consider a field component
\begin{equation}
U\left(  z'\right)
=\lambda^{-1}\int_{\sigma}d\mu~a_{\mu}Y_{\mu}\left(
z'\right)  ,\label{un}%
\end{equation}
where the integration is over the fuzzy segment $\sigma$. It is
natural to interpret this component as a contribution to the total
field of the beam of rays that form the selected segment. Using
(\ref{u-Y}) and (\ref{a-Y}), this expression can be rewritten as
\cite{V2017}

\begin{equation}
U\left(  z'\right)  =\int dz''~\Pi\left( z',z''\right)
u\left(z''\right)  , \label{Un-Q}%
\end{equation}
where
\begin{equation}
\Pi\left(  z',z''\right)
=\lambda^{-1}\int_{\sigma}d\mu~Y_{\mu}\left(
z'\right)  Y_{\mu}^{\ast}\left(  z''\right)  . \label{Q-Y}%
\end{equation}
Thus, the isolation of the beam contribution to the total field is
carried out using linear spatial filtering. By selecting the area
$\sigma$ the filter is 'tuned'  to the required beam.

In this paper, we consider beams that (i) are formed by rays with
the same identifier and (ii) cover the entire vertical section of
the waveguide at the observation distance. Examples of such beams
are shown in Fig. 5. However, the described procedure is
applicable to any beam of rays. In the case of free space, when
the field on the antenna is a fragment of a plane wave formed by a
beam of parallel rays, the described procedure is reduced to the
standard formation of an antenna lobe.

Isolation of the beam contribution by this method is possible only
if the fuzzy segments of the phase plane corresponding to
different beams do not overlap. This requirement can only be met
at sufficiently high frequencies. In Sec. \ref{sec:experiment} it
is noted that when processing experimental data, we can analyze
signals only in the 600$\pm$300 Hz band. Gray areas in the upper
panels of Fig. 6 and 7 represent fuzzy segments at 600 Hz, for -6
and +8 segments, respectively. In both cases, neighboring segments
fall into the gray areas. This means that the selected fuzzy
segments overlap with neighboring fuzzy segments, and when
processing the tonal signals, the contributions of corresponding
beams of rays cannot be resolved. However, we will see below that,
when dealing with pulsed signals, the spatial resolution is
supplemented by the temporal one, and the isolation of individual
beams becomes possible.

\subsection{Pulse signals \label{sub:pulses}}

Using (\ref{u-Y}) and (\ref{a-Y}), we represent the complex
amplitude $u\left( z,f\right) $ at each frequency $f$ as a
superposition of coherent states. Then the coherent state
amplitude $a_{\mu}$ becomes a function of $f$. When calculating
this function, the scale $\Delta_{z}$ can be chosen different for
different frequencies.

Let us introduce the function
\begin{equation}
a\left( z,p,t\right) =\int df~a_{\mu}\left( f\right) e^{-2\pi
ift},\label{a-pzt}%
\end{equation}
characterizing the distribution of the transient field amplitude
in phase space 'depth --momentum (angle) -- time' $\left(
z,p,t\right) $. For points of this space lying on a ray line, the
function $a\left( z,p,t\right) $ can be interpreted as the complex
amplitude of a sound pulse that comes to depth $z$ at grazing
angle $\arcsin p$ at time $t$.

Further, when processing the experimental data, the main attention
will be paid to the  analysis of the coherent state intensity
\begin{equation}
J\left( z,p,t\right)  =\left\vert a\left(  z,p,t\right)
\right\vert
^{2}.\label{J-def}%
\end{equation}
This function takes the largest values near the ray line, and
decreases as it moves away from it.

In Sec. \ref{sub:CS}, it is shown that in the case of a tonal
sound field, the distribution of coherent states intensity
$\left\vert a_{\mu}\right\vert^{2}$ is localized inside the fuzzy
ray line in the phase plane. In 3D phase space
$\left(z,p,t\right)$, we introduce a similar area, which will also
be called a fuzzy ray line. We define its boundary as follows. The
signal arrival times $t$ at the observation distance belong to a
certain interval $t_{\min}<t<t_{\max}$. Let's choose an arbitrary
time $t_{\ast}$ from this interval and consider the plane $\left(
z,p\right) $ formed by points of the phase space with
$t=t_{\ast}$. This plane intersects the ray line at, generally
speaking, several points. Timefronts in the middle panels of Figs.
6 and 7 show that in our waveguide model, for any $t_{\ast}$,
there are two intersection points. Examples of such pairs are
points A and A' in Fig. 6 and points B and B' in Fig. 7. Let
$\mu_{\ast}=\left( z_{\ast},p_{\ast}\right) $ be one of the
intersection points. As the boundary of the fuzzy ray line in the
neighborhood of $\mu_{\ast}$, we take the ellipse formed by the
points $\mu$ that are spaced from $\mu_{\ast}$ by the
dimensionless distance $d\left( \mu,\mu_{\ast}\right) =1$. The
dimensionless distance (\ref{d}) depends on the frequency $f$, and
when defining the boundary, we evaluate this distance for the
center frequency of the analyzed signal. In our case, this is $f$
= 600 Hz. Figure 8 presents fragments of fuzzy ray lines at
distances of 380 (top) and 905 m (bottom). The segments of the ray
line shown in Fig. 4, here turned into tubes of finite thickness.

\begin{figure}[!t]
\centering
\includegraphics[width=4.5in]{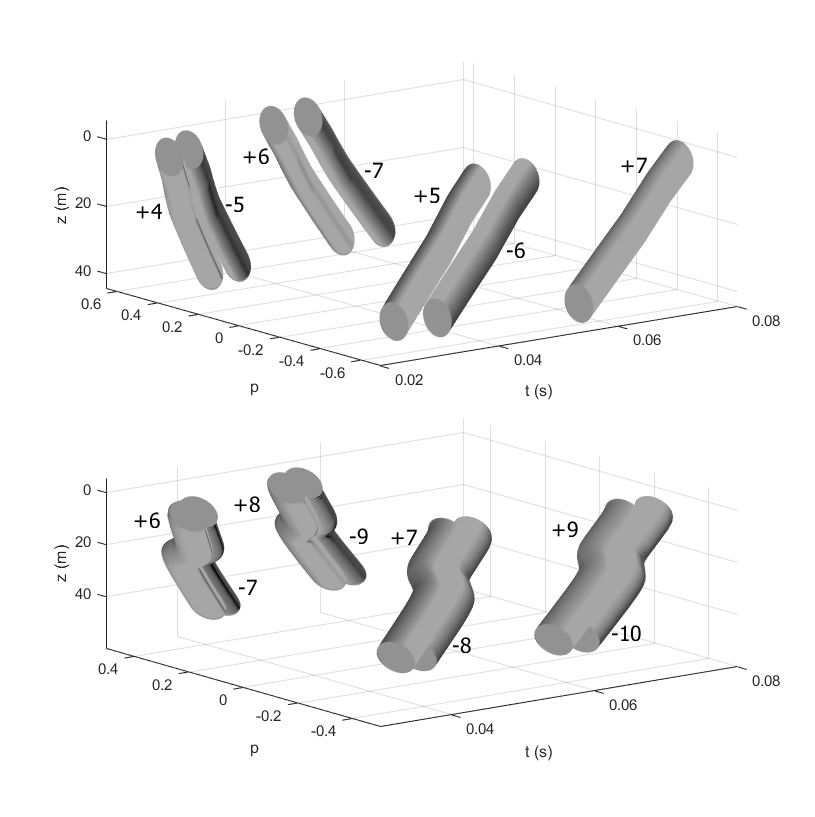}
\caption{Fragments of fuzzy ray lines at distances of 380 (top)
and 905 m (bottom). The corresponding identifier is indicated next
to each fuzzy segment.} \label{fig_8}
\end{figure}

The procedure for isolating  the ray beam contribution to the tonal field, 
described in Sec. \ref{sub:isolation},  is naturally generalized to the case of a 
pulsed source. First, the procedure is used to isolate the beam contribution $U\left(z,f\right)$ from the component  of the total field $u\left(z,f\right) $ at 
each frequency $f$ in the band of the emitted signal. Then the beam contribution to 
the total field $v\left( z,t\right) $ is synthesized via the Fourier 
transformation%

\begin{equation}
V\left(  z,t\right)  =\int df~U\left(  z,f\right)  e^{-2\pi ift}. \label{Vzt}%
\end{equation}

Analytical relations, presented above, suggest, that the amplitude
of the coherent state $a_{\mu}$ associated with a point $\mu =
(z,p)$ in the phase space  is determined mainly by the
contributions of rays arriving in the depth interval
$z\pm\Delta_z/2$. In the examples considered below, such rays have
approximately the same bottom reflection coefficients $V$. As
indicated in Sec. \ref{sec:experiment}, in this paper we consider
the field components that are completely reflected from the bottom
and therefore $V = \exp(i\phi)$. Due to the inaccuracy of the
bottom model, we cannot correctly calculate the coefficient $V$
and therefore the complex amplitude $a_{\mu}$ can be found only up
to an unknown phase factor. However, the intensity $|a_{\mu}|^2$
is calculated correctly in this case. Moreover, we assume a weak
dependence of the coefficient $V$ on the frequency (in the band of
the emitted signal). Then the intensity of the pulsed field
$J\left(z,p,t\right)$ is also correctly predicted. The validity of
these assumptions is confirmed by the comparison of theory and
experiment presented in the next section.

\section{Data analysis \label{sec:data}}

When processing experimental data, our attention was focused on
calculating the distribution of the sound field intensity $J\left(
z,p,t\right) $ in the 3D phase space $\left( z,p,t\right) $. The
results of calculating these distributions for fields recorded at
distances of 380 and 905 m are compared with theoretical
predictions. Similar comparisons made for the other three
distances look similarly and are therefore not presented here.
This section also demonstrates that the obtained intensity
distributions $J\left(z,p,t\right) $ can be used to estimate the
distance to the source. This is done for the all five distances.

One of the reasons for the discrepancy between theory and
experiment is the errors in the reconstruction of the field in the
vertical section of the waveguide from the measurement data at
only 10 horizons. To assess the influence of this factor, we
present two variants of numerical simulation. The first of them is
performed for a receiving array of a large number of elements
(with an interelement distance small compared to the wavelength),
which uniformly fill the entire vertical section of the waveguide.
The second variant is based on the theoretical calculation of the
sound field at ten points of the vertical section corresponding to
the hydrophone depths. The field in the entire vertical section is
reconstructed in exactly the same way as in the processing of
experimental data. On the plots representing the results obtained
by the first and second methods, it is indicated  'Theory, dense
array' and 'Theory, sparse array', respectively.

In the coherent state expansion for all distances and all
frequencies, the vertical scale $\Delta_{z}$ is equal to $10$ m.

\subsection{Intensity distribution in phase space ($z,p,t$)
\label{sub:J}}

Let us consider the function $J\left(z,p,t\right) $ representing
the intensity distribution of the field registered at a distance
of 380 m. A fragment of the ray line in 3D phase space for this
distance is shown in the upper panel of Fig. 4, and the
projections of this line on the planes $\left( p,z\right) $,
$\left( t,z\right) $, and $\left( t,p\right) $ are in Fig. 6. The
point representing the arrival of ray $A$ has the coordinates
$\left( z_{A},p_{A},t_{A}\right) $, where $z_{A}$ = 23.4 m,
$p_{A}$ = -0.52, and $t_{A}=$0.4 s.

Figure 9 presents a section of the function $J\left( z,p,t\right)
$ by the plane $t=t_{A}$. At points $A$ and $A^{\prime}$, shown by
white circles, this plane is crossed by ray line segments -6 and
+5, respectively. In the top panel of Fig. 9 we see that the
points $A$ and $A^{\prime}$ are located near the local maxima of
the intensity distribution in the plane $\left( p,z\right) $. Note
that the point $A^{\prime}$ is located on the horizon 3.9 m,
located outside the depth interval covered by the receiving array.
Therefore, the field near $A^{\prime}$ is poorly reconstructed,
and the corresponding local maximum of the intensity distribution
is hardly noticeable on the middle and lower panels. Magenta
ellipses in the upper panel are formed by points that are spaced
from A and A' by dimensionless distances $d=1$ (at a frequency of
600 Hz). These ellipses represent the intersections of the
$t=t_{A}$ plane and the boundaries of the fuzzy segments
corresponding to identifiers -6 and +5.

\begin{figure}[!t]
\centering
\includegraphics[width=4.5in]{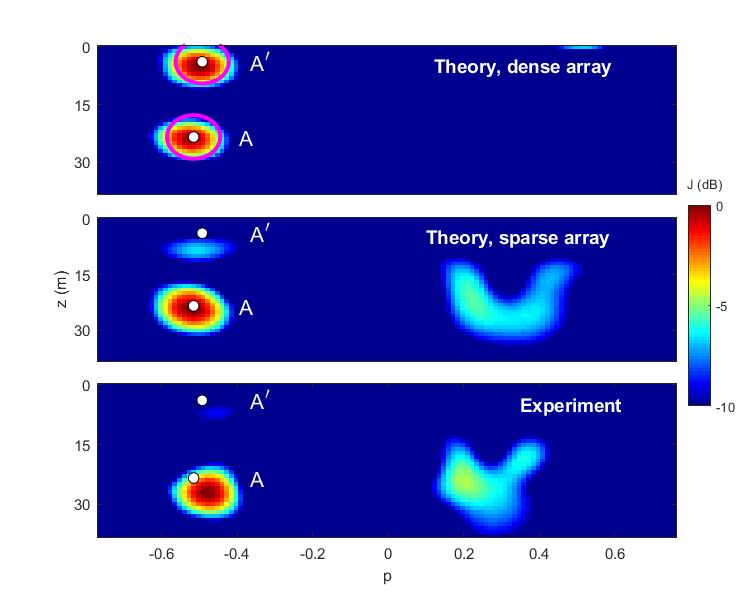}
\caption{Cross section of the intensity distribution $J\left(
z,p,t\right) $ at a distance of 380 m by the plane $t=t_{A}$. Top
and middle panels: numerical simulations. Bottom panel:
experiment. The white circles indicate the arrivals of rays A and
A' in the phase plane. Magenta ellipses in the upper panel show
the boundaries of fuzzy segments +6 (around point A) and -5
(around point A') in the plane $t=t_{A}$.} \label{fig_9}
\end{figure}

Figure 10 shows a section of the function $J\left( z,p,t\right) $
by the plane $z=z_{A}$. This plane intersects all segments of the
ray line, and in the vicinity of each intersection point, an area
of increased intensity is observed. The identifier of the
corresponding segment is indicated next to each area of high
intensity. Horizon $z_{A}$ is located within the depth interval
covered by the receiving array. Therefore, the field in the
vicinity of this point is reconstructed quite well, and the
intensity peaks predicted by the theory are clearly distinguished
on the all three panels.

\begin{figure}[!t]
\centering
\includegraphics[width=4.5in]{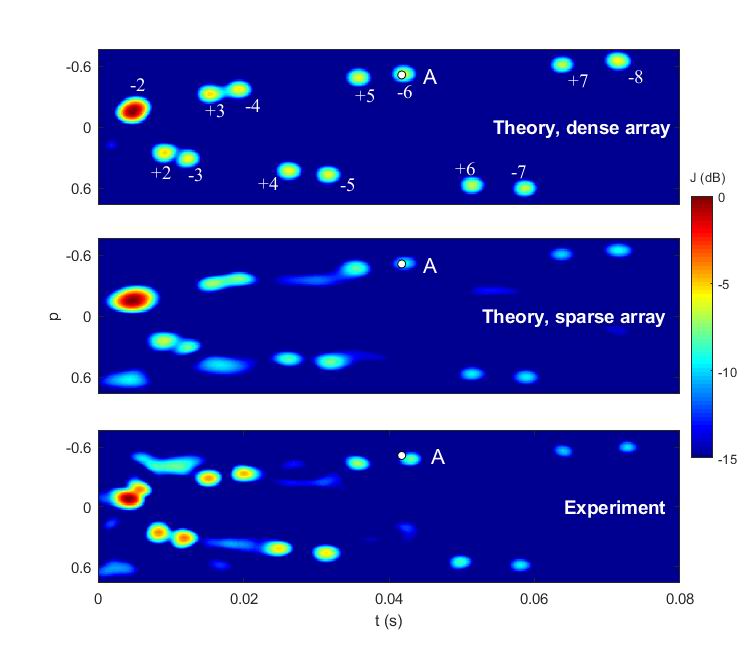}
\caption{Cross section of the intensity distribution $J\left(
z,p,t\right) $ at a distance of 380 m by the plane $z=z_{A}$. Top
and middle panels: numerical simulations. Bottom panel:
experiment. The white circles indicate the arrival of ray $A$. In
the upper panel, next to each local maximum, the identifier of the
corresponding segment is indicated.} \label{fig_10}
\end{figure}

The section by the plane $p=p_{A}$ is shown in Fig. 11. This plane
intersects the ray line segments -6 and +5. The projections of
these segments onto the $\left( t,z\right) $ plane are shown as
white dotted lines. The point of the intersection with segment -6,
marked with a white circle, depicts the arrival of ray $A$. The
rays forming segments -6 and +5 at the observation distance have
momenta $p$ close to $p_{A}$. Therefore, in this particular
example, the obtained intensity distribution in the time-depth
plane $\left( t,z\right) $ practically coincides with the
intensity distribution $\left\vert V\left( z,t\right) \right\vert
^ {2}$ at the output of the filter, determined by equations
(\ref{Un-Q}) and (\ref{Vzt}), and tuned to isolate a beam of rays
with identifier -6. We have already noted (Sec.
\ref{sub:isolation}) that the fuzzy segment -6 shown in the top
panel of Fig. 5 overlaps with segment +5. This means that when
dealing with tonal field at a frequency of 600 Hz, the
contributions of ray beams with identifiers +5 and -6 cannot be
separated. In Fig. 10 we see that in the case of a pulsed source,
the additional temporal resolution made it possible to isolate the
contributions of these beams.

\begin{figure}[!t]
\centering
\includegraphics[width=4.5in]{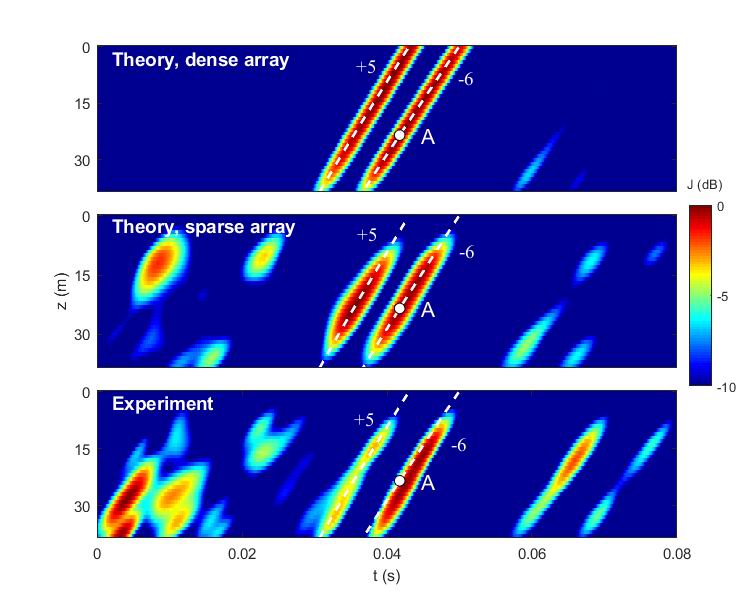}
\caption{Cross section of the intensity distribution $J\left(
z,p,t\right) $ at a distance of 380 m by the plane $p=p_{A}$. Top
and middle panels: numerical simulations. Bottom panel:
experiment. The white circle shows the arrival of ray $A$. The
white dotted lines show the projections of segments +5 and -6.}
\label{fig_11}
\end{figure}

Similar results for a distance of 905 m are shown in Figs. 12-14.
In these figures we see the sections of the distribution $J\left(
z,p,t\right) $  by planes passing through the point $\left(
z_{B},p_{B},t_{B}\right)$ representing the arrival of ray $B$,
where $z_{B}$ = 27.3 m, $p_{B}$ = 0.4, $t_{B}=$ 0.05 s.  Figures
12, 13, and 14 present the sections by the planes $t=t_{B}$,
$z=z_{B}$, and $p=p_{B}$, respectively. The white circles show the
arrivals of rays $B$ and $B^{\prime}$. Points B and B' belong to
segments +8 and -9, respectively. Therefore, the magenta ellipses
in Fig. 12 and white dotted lines in Fig. 14 are built for these
segments. In the top panel of Fig. 13 we see that, unlike the
results obtained for the distance of 380 m (Fig. 10), even when
using a dense array covering the entire vertical section, the
contributions of not all beams are resolved in the $z=z_{B}$
plane.

The results presented in Figs. 9-14, as well as similar results
obtained for other three distances, show that the intensity
distribution $J(z,p,t)$, even in the absence of complete
information about the bottom parameters, can be satisfactory
predicted using the simplest range-independent environmental
models.

\begin{figure}[!t]
\centering
\includegraphics[width=4.5in]{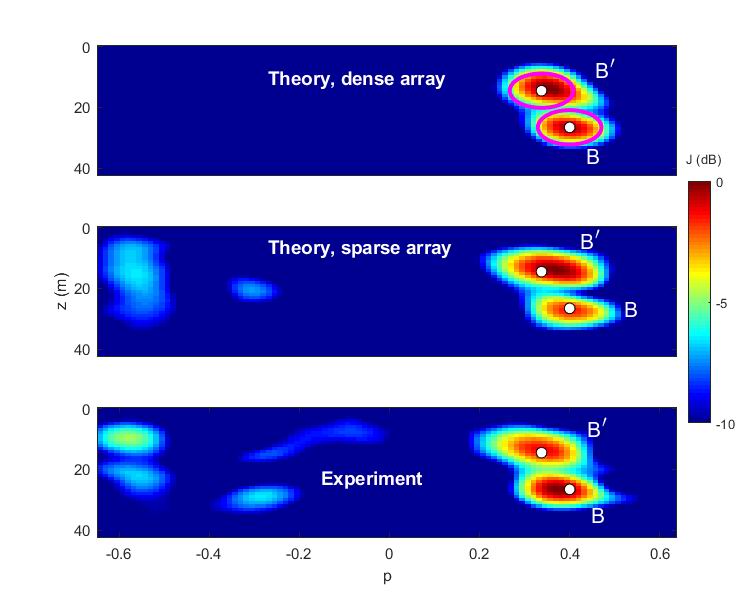}
\caption{Cross section of the intensity distribution $J\left(
z,p,t\right) $ at a distance of 905 m by the plane $t=t_{B}$. Top
and middle panels: numerical simulations. Bottom panel:
experiment. The white circles indicate the arrivals of rays B and
B' in the phase plane. Magenta ellipses in the upper panel show
the boundaries of fuzzy segments +8 (around point B) and -9
(around point B') in the plane $t=t_{B}$.} \label{fig_12}
\end{figure}

\begin{figure}[!t]
\centering
\includegraphics[width=4.5in]{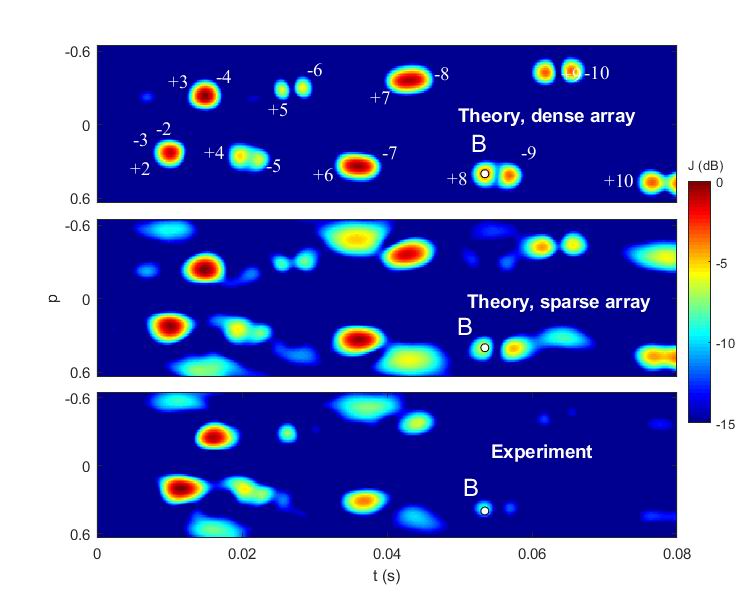}
\caption{Cross section of the intensity distribution $J\left(
z,p,t\right) $ at a distance of 905 m by the plane $z=z_{B}$. Top
and middle panels: numerical simulations. Bottom panel:
experiment. The white circles indicate the arrival of ray $B$. In
the upper panel, next to each local maximum, the identifier of the
corresponding segment is indicated.} \label{fig_13}
\end{figure}

\begin{figure}[!t]
\centering
\includegraphics[width=4.5in]{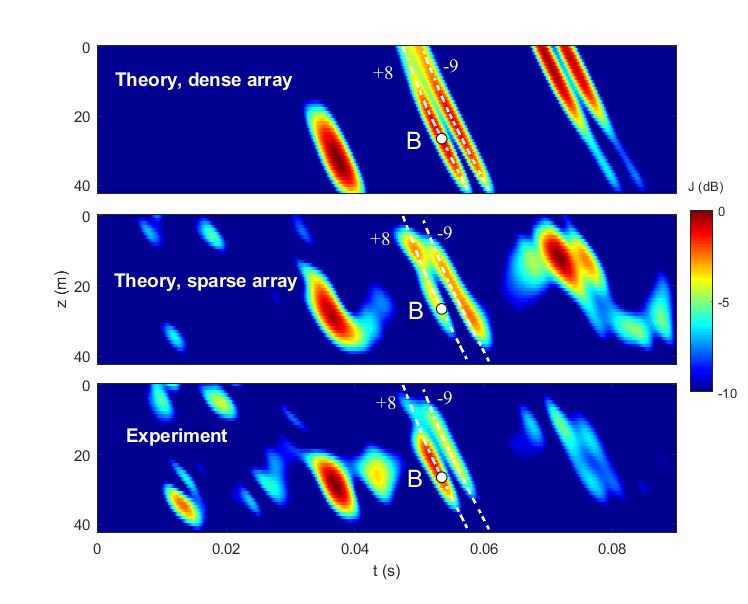}
\caption{Cross section of the intensity distribution $J\left(
z,p,t\right) $ at a distance of 905 m by the plane $p=p_{B}$. Top
and middle panels: numerical simulations. Bottom panel:
experiment. The white circle shows the arrival of ray $B$. The
white dotted lines show the projections of segments +8 and -9.}
\label{fig_14}
\end{figure}

\subsection{Source localization \label{sub:localization}}

Comparison of theory and experiment confirms our assumption that
the intensity distribution in the phase space is localized mainly
inside the fuzzy ray line introduced in Sec. \ref{sub:pulses}.
This circumstance can be used to estimate the coordinates of the
source, that is, to solve the localization problem.

Our idea is as follows. Consider a set of possible source
positions $\mathbf{r}$ forming a search grid. Using the standard
ray code, for each position $\mathbf{r}$, we find all the beams of
rays escaping this point and hitting the antenna. Based on this
calculation, we construct the function $W\left(
z,p,t;\mathbf{r}\right) $, which is equal to 1 if the point
$(z,p,t)$ is inside the fuzzy ray line, and 0 otherwise.
Introduce the uncertainty function%
\begin{equation}
K\left(  \mathbf{r}\right)  =\max_{\tau}\int
dpdzdt~J\left(z,p,t\right)
W\left(  z,p,t+\tau;\mathbf{r}\right)  , \label{K}%
\end{equation}
where $J\left( z,p,t\right) $ is the intensity distribution of the
registered field. It is natural to expect that the function
$K\left( \mathbf{r}\right) $ will take its maximum value at the
point $\mathbf{r}$, which coincides with the actual source
position $\mathbf{r}_{s }$. In this case, the weight function
$W\left( z,p,t+\tau;\mathbf{r}\right) $ provides integration over
those areas of the phase space where $J$ takes the largest values.
A search by $\tau$ is necessary, since it is assumed that we do
not know the exact time of signal emission. The desired estimate
of the source position has the form
\[
\mathbf{\hat{r}}_{s}=\arg\max_{\mathbf{r}}K\left(
\mathbf{r}\right)  .
\]

Let us apply this approach to estimate the source coordinates from
the data of acoustic measurements. In our range-independent
waveguide model, the uncertainty function (\ref{K}) has two
arguments $r$ and $z$. Function $J\left(z,p,t\right)$ in the
integrand in (\ref{K}) is calculated by the coherent state
expansion of field $u(z,t)$ in the vertical section of the
waveguide at the location of the receiving array. Figure 15 shows
the uncertainty functions $K\left( r,z\right)$, obtained using the
fields $u(z,t)$ calculated theoretically on a dense vertical array
covering the entire vertical section of the waveguide at 340 m
(a), 380 m (b), 415 m (c), 905 m (d), and 1440 m (e). Similar
calculations of the uncertainty functions were carried out using
the fields $u(z,t)$ reconstructed from the signals recorded by 10
hydrophones. Figures 16 and 17 show the results obtained for
hydrophone signals calculated theoretically and measured in
experiments, respectively. It is seen that in all cases the white
asterisk, indicating the point of the actual source position, is
located inside the main peak of the uncertainty function $K\left(
r,z\right)$.

\begin{figure}[!t]
\centering
\includegraphics[width=4.5in]{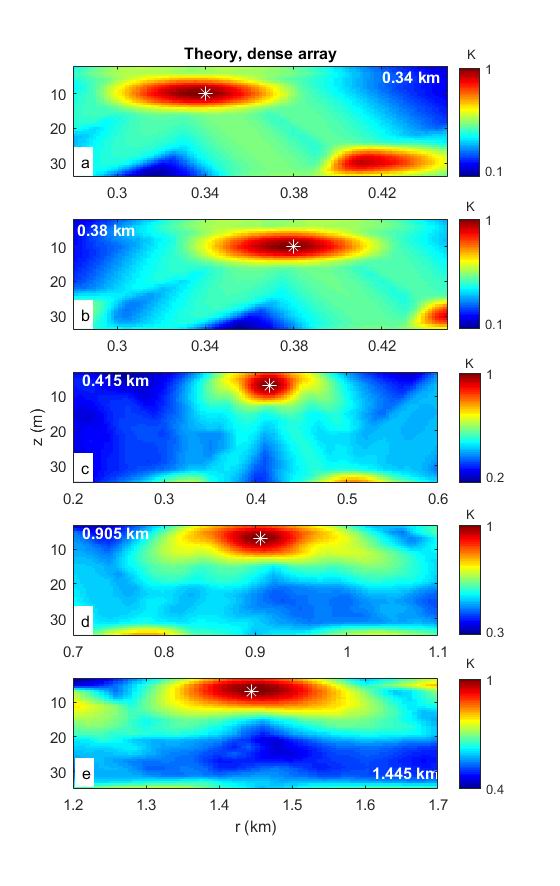}
\caption{Uncertainty functions $K\left( r,z\right) $ for the
observation  distances of 340 m (a), 380 m (b), 415 m (c), 905 m
(d) and 1440 m (e). The functions are obtained using the
theoretical calculation of the field on a dense receiving array
covering the entire vertical section of the waveguide. The white
asterisk indicates the actual source position.} \label{fig_15}
\end{figure}

\begin{figure}[!t]
\centering
\includegraphics[width=4.5in]{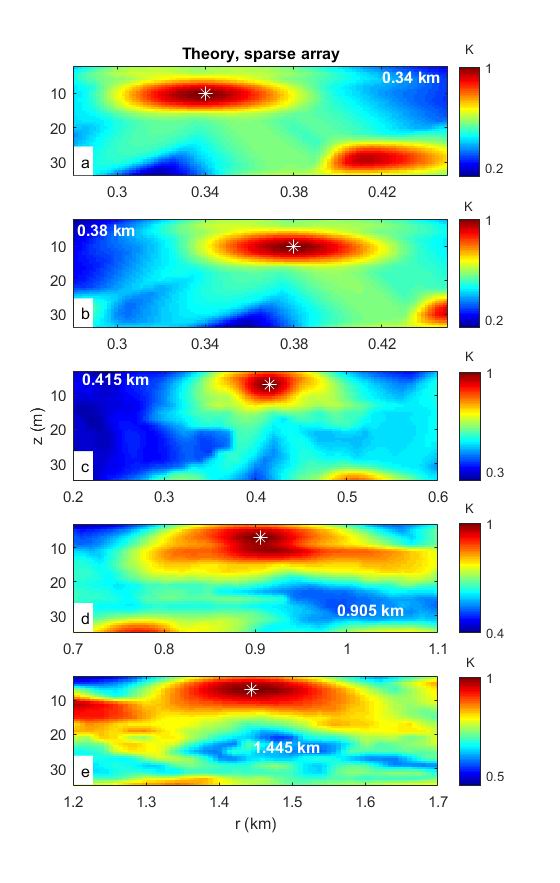}
\caption{The same as in Fig. 15, but for a receiving array of 10
elements.} \label{fig_16}
\end{figure}

\begin{figure}[!t]
\centering
\includegraphics[width=4.5in]{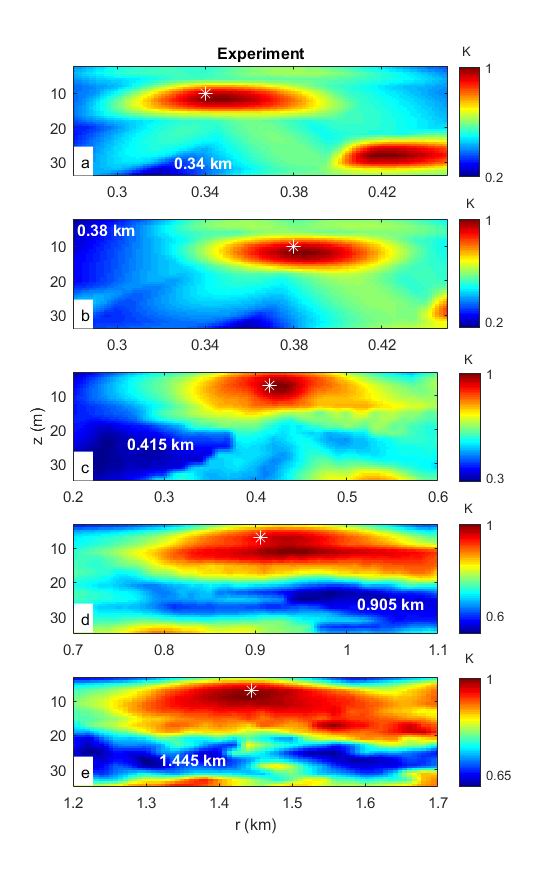}
\caption{The same as in Fig. 15, but with the uncertainty
functions computed using the experimental data.} \label{fig_17}
\end{figure}

\section{Conclusion \label{sec:conclusion}}

This work continues the analysis of the phase space representation
of the sound field in an underwater waveguide started in Refs.
\cite{V2017,V2020a}. Such a description of the wave field is
unconventional for underwater acoustics. Our interest in this
representation is due to the fact that the field distribution in
the 3D  phase space 'depth - angle - time' is more regular and
predictable than in the 2D space 'depth - time'. The point is that
there are no multipaths and no problems with caustics in the phase
space \cite{Alonso2010}. Refs. \cite{V2017,V2020a} argue that when
solving inverse problems, the transition from the configuration
space to the phase space makes it possible to relax the
requirements for the accuracy of the environmental model.

The phase space representation of the sound field in the vertical
section of the waveguide is introduced using the coherent state
expansion. This allows one to resolve the contributions of waves
simultaneously in terms of depth and arrival angle, that is, to
isolate signals arriving in a relatively small depth interval at
grazing angles from a relatively small angular interval. The
considered approach makes it possible to find the distributions of
the field amplitude and intensity in the phase space. Explicit
expressions for calculating these distributions are given by
(\ref{a-Y}) and (\ref{a-pzt}).

Our attention was mainly focused on the analysis of the intensity
distribution in the phase space $J\left( z,p,t\right) $. Since the
sound field was recorded using the array of only ten sparsely
spaced elements, in the processing we analyzed only the field
components at frequencies in the lower part of the emitted signal
band. Comparison of the results presented in the upper and middle
panels of Figs. 9-14 shows that the undersampling, even at low
frequencies, causes some errors in the intensity calculation.
However, there is good agreement between theory and experiment.
Since the theoretical description is based on a highly idealized
environmental model, this fact confirms our assumption that the
field intensity distribution in the phase space is weakly
sensitive to variations in the waveguide parameters.

An important implication of the theory is that the intensity
distribution is localized in the area of the phase space, which we
call the fuzzy ray line. Figure 8 show that these area occupy a
relatively small part of the phase space available for rays in our
waveguide model. In Sec. \ref{sub:localization} it is shown that
this circumstance can be used to solve the problem of source
localization in a waveguide.

Finally, we note one more important application of the coherent
state expansion. Section \ref{sub:isolation} shows that it can be
used to isolate a component of the total field  formed by a given
beam of rays. In the present paper there is no separate section
describing the application of this procedure, generalizing the
standard procedure for forming a lobe of directivity pattern in
free space. The point is that in our example, isolating the
contribution from the ray beam is actually equivalent to
calculating the cross section of the intensity distribution $
J\left( z,p,t\right) $ by the plane $p=$ const. In the cross
sections presented in Figs. 11 and 14 we clearly see the isolated
contributions of beams of rays with identifiers +5 and -6 at a
distance of 380 m and with identifiers +8 and -9 at a distance of
905 m, respectively.

\section*{Acknowledgment}
Authors are grateful to Dr. A. Lunkov for the help in experimental
data processing. The research was carried out within the state assignment of Institute of Applied Physics of the Russian Academy of Sciences (Project 0030-2021-0018). It was also supported in part by Israel Science Foundation, grant 946/20. 


\end{document}